\documentstyle[prl,multicol,aps,epsfig]{revtex}

\begin{document}

\title{
Quasiperiodic patterns in boundary-modulated excitable waves}
\author{Irene Sendi\~{n}a-Nadal$^{1}$, Vicente P\'{e}rez-Mu\~{n}uzuri$^{1}$,
V\'{\i}ctor M. Egu\'{\i}luz$^{2}$, Emilio Hern\'{a}ndez-Garc\'{\i}a$^{3}$ and Oreste
Piro$^{3}$
}
\address{
$^{1}$ Grupo de F\'{\i}sica non Lineal, Universidade de Santiago
de Compostela. E-15706 Santiago de Compostela, Spain.
\\$^{2}$
Centre for Chaos and Turbulence Studies, Niels Bohr Institute,
Blegdamsvej 17, 2100 Copenhagen Ø,
Denmark.
\\$^{3}$ Instituto Mediterr\'aneo de Estudios Avanzados
(IMEDEA), CSIC-Universitat de les Illes Balears.
\\E-07071 Palma de Mallorca, Spain. }

\date{\today}
\maketitle

\begin{abstract}
We investigate the impact of the domain shape on wave propagation
in excitable media.
Channelled domains with sinusoidal boundaries are considered.
Trains of fronts generated
periodically at an extreme of the channel are found to adopt a
quasiperiodic spatial configuration stroboscopically frozen in time.
The phenomenon is studied in a model for the photo-sensitive
Belousov-Zabotinsky reaction, but we give a theoretical
derivation of the spatial return maps prescribing
the height and position of
the successive fronts that is valid for arbitrary excitable
reaction-diffusion systems.

\end{abstract}

\begin{multicols}{2}
\narrowtext
Excitable media display a very rich spatio-temporal behavior with
regimes ranging from fairly well ordered structures of propagating
waves \cite{kapral} to highly uncorrelated spatio-temporal chaos.
The study of all these features as well as their mutual
connections provides very useful insight to understand and
eventually control phenomena of paramount applied importance such
as the deadly arisal of fibrillation in cardiac tissues\cite{PNAS}
or the appearance of either ordered or turbulent patterns in
extended chemical reactors operating away from equilibrium
conditions. In many of these applications, a crucial but
frequently ignored ingredient is the presence of boundaries.
For example, it has been shown that boundaries and obstacles in
inhomogeneous media are important to either pin or repele spiral patterns
\cite{moncho_davidenko_sepulcre_aranson}; moving boundaries \cite{alberto}
and stripped configurations \cite{zhabo,steinbock} 
have also strong effects.

Unfortunately, the current understanding 
of boundary effects in
nonlinear partial differential equations is rather
incomplete, and sometimes surprisingly nontrivial behavior lurk
behind the apparent simplicity of some problems. A recent
study\cite{eguiluz}, for example, shows that relatively
regular boundary conditions such as Dirichlet's on the
banks of a sausage-shaped channel can elicit several types
of spatial complexity such as frozen quasiperiodicity and chaos
even in very simple reaction diffusion equations.
There, the axial coordinate along the channel acts as a ``time"
in the equations describing the
time-independent spatial patterns and the undulated
boundaries play the role of a periodic force inducing
chaos in a dynamical system that is non-chaotic in the absence of
driving.

On the other hand, propagation of waves in excitable media has
been studied in a variety of contexts\cite{kapral}. Due to their
ubiquity in large two-dimensional systems, much of this work deals
with spiral waves and focuses, in particular, on the various
aspects of the behavior around the cores of the spiral patterns.
In contrast, the propagation of front trains has received much
less attention. This may seem surprising since the same spirals
can be seen far from their cores as a periodic train of
two-dimensional traveling fronts. These trains, though, are easily
characterized by a dispersion relation $c=c(\lambda)$, giving a
relation between the constant front train velocity and its uniform
spacing $\lambda$ and their dynamics is very simple.
However, much less trivial behavior appears even in one-dimensional
systems if the excitable medium recovers the
rest state not monotonically but via damped oscillations\cite{meron}.
In this regime, propagating wave trains often relax
to irregularly spaced configurations of fronts that can be seen as
spatial chaos. 

The purpose of this Letter is to report a new kind of nontrivial
spatial structure arising as a pure boundary effect in excitable media,
namely, stroboscopically-frozen
quasiperiodicity. We investigate the asymptotic propagation of
excitable wave trains generated by local time-periodic stimulation
at the extreme of a sinusoidally undulated channel. We
find that contrarily to intuitive expectations, the trains of
fronts asymptotically accommodate in quasiperiodic spatial
configurations, incommensurated with the boundaries but periodic in
time and synchronized with the stimuli. With the experiments on
the photosensitive Belousov-Zhabotinsky reaction in mind
\cite{alberto,steinbock}, we demonstrate this
phenomenon in the Oregonator model adapted to include
the effect of light. Finally, we present a more general semi-analytic
theory of the formation of the quasi-periodic, and
possibly chaotic, structures referred above.

Photosensitive $Ru(bby)_{3}^{+2}$-catalyzed
Belousov-Zhabotinsky reactive media can be modelled \cite{krug}
by the following version of the Oregonator model:

\begin{eqnarray}
\frac{\partial\,u}{\partial\,t} & = & \frac{1}{\varepsilon}
\left(u\,-\,u^{2}\,-\,(f\,v\,+\,\phi)\,\frac{u\,-\,q}{u\,+\,q}\right)\,+\,D_{u}\,\nabla^{2}u
\nonumber\\
\frac{\partial\,v}{\partial\,t} & = & (u\,-\,v)\ .
\label{eq1}
\end{eqnarray}
\noindent
Here $u$ (resp. $v$) describe $HBrO_{2}$ (resp. catalyst)
concentrations. $D_{u}$ is a diffusion coefficient and $f$, $q$ ,  $\varepsilon$
and $\phi$ are
parameters related to the reaction kinetics.
In our simulations we set 
$f=3$, $q=0.002$, $\varepsilon=0.05$, $\phi=0.002$ and $D_u=1$.

We simulate this reaction in a spatial domain tailored as
an undulated channel
along the longitudinal direction $x$. The transversal coordinate
$y$ is bounded by two sinusoidal walls of common spatial frequency
$k = 2\pi / \lambda _p$, amplitude $d$, minimum separation $s$,
and phase mismatch $\eta$:

\begin{eqnarray}
y_0(x) & = & \frac{d}{2}[1-\cos (k x)] \nonumber \\
y_1(x) & = & -s -\frac{d}{2}[1+\cos (k x + \eta)]
\label{bound}
\end{eqnarray}

\begin{center}
\begin{figure}
\epsfig{file=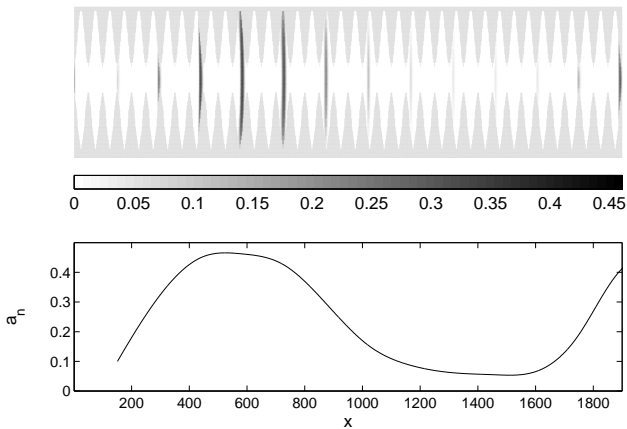, height=5.7cm}
\vspace{0.1in}
\caption[]{In the upper panel, the white area limited by gray
undulated boundaries is the excitable region where fronts
originated at the left end propagate. The transverse features
are the fronts travelling to the right. Darker fronts have a
larger value of $u$, as indicated in the colorbar. The lower panel
displays the envelope of the maximum amplitude of the fronts.}
\label{fig:foto_canal}
\end{figure}
\end{center}

\noindent
We will concentrate on symmetric
sausage-shaped channels ($\eta = \pi $) of width $w(x)=s+d- d
\cos(k x)$, as the one shown in Fig.~\ref{fig:foto_canal}.
On the sinusoidal boundaries we impose the Dirichlet condition
$u=0.004$, a value close
to the fixed point of the local dynamics.

In order to solve numerically Eq.~(\ref{eq1}) it is convenient to map
the region limited by $y_{0}(x)$ and $y_1(x)$ and by $x=0,L$ to a
rectangle: $\widetilde{y} _{1}=1$, $\widetilde {y}_0=0$, and
$x=0,L$, where $\widetilde y = (y-y_0)/(y_1-y_0)$
and $L$ is the length of the channel.
Under this map, the diffusion term transforms as\cite{eguiluz}:

\begin{equation}
\nabla ^2 u \rightarrow \partial ^2_{xx}\widetilde u +
F(x)\partial^2_{\widetilde y \widetilde y}
\widetilde u+ G(x)\partial ^2_{x \widetilde y} \widetilde u +
H(x) \partial_{\widetilde y} \widetilde u \ ,
\label{lapla}
\end{equation}
$F(x)$, $G(x)$ and $H(x)$, given in \cite{eguiluz}, are periodic
functions reflecting the undulations of the boundaries via
modulations measured by the product $ k d $. In the limit $ k d \rightarrow 0 $
(straight channel),
Eq.(\ref{lapla}) becomes the standard Laplacian.

Wave trains are generated stimulating the medium at the
left end $x=0$ of the channel by
pushing $u$ above and below the
excitability threshold periodically in time.
The opposite end of the channel is set as a no-flux boundary.
During the simulations we
mainly varied the forcing parameters $\lambda_p$ and $d$, but
also several wave train periods and channel widths
were investigated. After a transient, the fields $u$ and $v$ converge
to a configuration of propagating fronts that repeats itself periodically
in time in synchrony with the wave generator at $x=0$. In other words,
the train becomes a stroboscopically frozen pattern.
We denote by $x_n$ and $a_n$
the longitudinal position and maximum height of $u$ at the channel
axis, respectively, for the $n$-th front.

\begin{center}
\begin{figure}
\epsfig{file=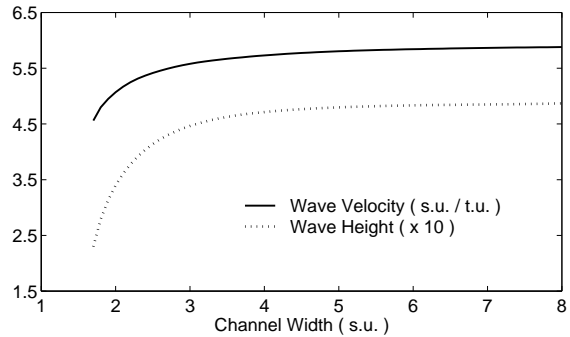, height=4.5cm}
\vspace{0.1in}
\caption[]{Wave train velocity and wave height (multiplied by a factor 10)
in straight channels of different widths. There is a critical
value, $s_c=1.65$, below which propagation becomes impossible.
$T=5$. }
\label{fig:canal_recto}
\end{figure}
\end{center}

As a comparison, in a straight channel
($d=0$) of finite width $s$ the asymptotic configuration of the wave
fronts is equally spaced by a length $\lambda$ and propagates with
velocity $c=\lambda/T$ if the forcing period is $T$. This velocity
increases with the channel width \cite{steinbock} starting from a
critical value $s_c$ of the latter, below which the fronts cannot propagate.
In Fig.~\ref{fig:canal_recto} we plot the train velocity and the
maximum amplitude of the wave fronts as a function of the width
of the straight channel.
In modulated domains with $d \neq 0$ a wide range of new
spatial configurations incommensurated with the boundaries emerge.
Typically, both the spacing and the amplitude of the fronts
become spatially quasiperiodic. According to the strength
$ k d \propto d / \lambda_p$, of the spatial forcing we
distinguish strong from weak modulations. Let us describe
the cases
$\lambda_p=50$ and $\lambda_p=1000$, respectively, as an
illustration.

\begin{center}
\begin{figure}
\epsfig{file=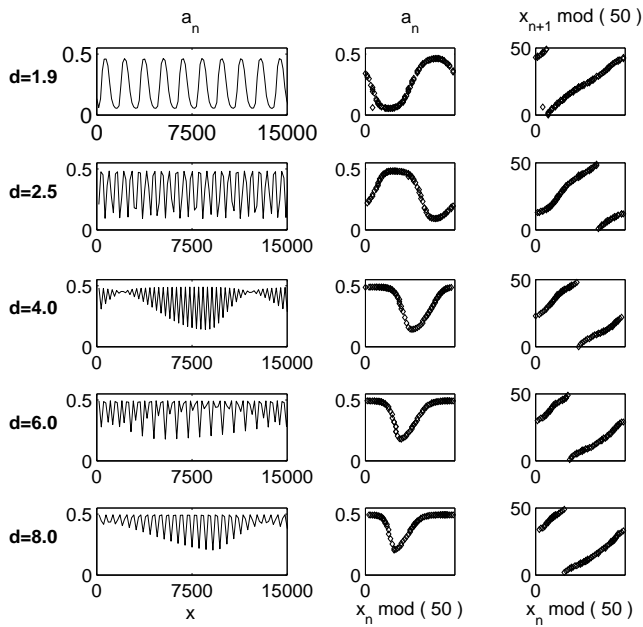, width=8.5cm}
\vspace{0.1in}
\caption[]{Numerical results obtained from Eq.~(\ref{eq1}) for
strong forcing. First column: Maximum height of each front within
the train, as a function of position $x$. Second column: Same as
before but with position $x$ folded modulo $\lambda_p=50$. Third
column: Return map of the front positions modulo $\lambda _p$.
Each row is for a different amplitude $d$. Parameters:
$\lambda_p=50$, $T=5$, $s=1.1$. Note that channel length $L=15000$
is much larger that in Fig.~\ref{fig:foto_canal}. }
\label{fig:TresxSete}
\end{figure}
\end{center}

\begin{center}
\begin{figure}
\epsfig{file=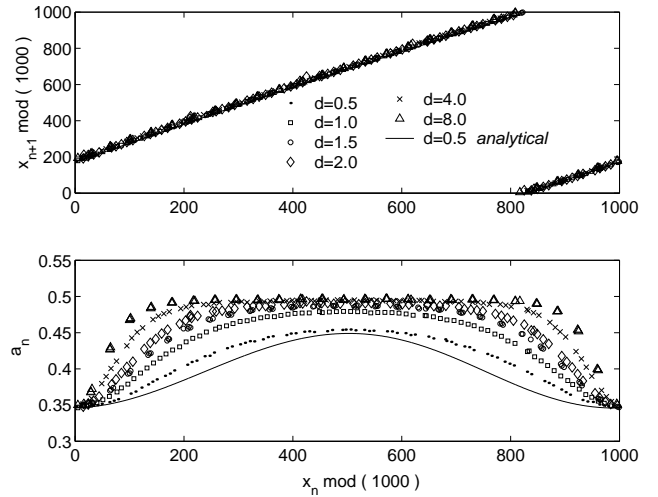, width=8.5cm}
\vspace{0.1in}
\caption[]{Weak forcing behavior at
$\lambda_p=1000$ and $s=2$. Above: Return map of the front
positions modulo $\lambda_p$. Below: Maximum height of each front
as a function of the front position modulo $\lambda_p$. Solid
lines are from Eqs.~(\ref{mapx}) and (\ref{mapalt}).}
\label{fig:xn_an}
\end{figure}
\end{center}

The results for strong modulation are shown in
Fig.~\ref{fig:TresxSete}. The amplitude of the boundary undulation
increases from top to bottom. The quasiperiodic behavior of
the pulse height becomes evident as $d$ increases.
The second column in Fig.~\ref{fig:TresxSete} shows also the
maximum $a_n$ of each front as a function of its position
modulo $\lambda_p$. This plot provides information
about the distribution of the front height maxima relative to the
elementary unit of the channel. 
Notice that the fronts do not always reach their minimal height at the
narrowest channel sections ($x=m\lambda_p$)
as one would naively expect from the behavior in straight
channels depicted in Fig.~\ref{fig:canal_recto}.
Moreover, the fronts can now propagate even when
the channel is narrower ($s=1.1$) in some places
than the minimun width $s_c=1.65$ that allows propagation in
straight channels.
Last column in Fig.~\ref{fig:TresxSete} displays the
{\it return maps} of the $(n+1)$-th front position $x_{n+1}$ (relative to the
unit channel cell) as a function of the position $x_n$ of the previous
front. The shapes of the curves are analog
to those {\it circle maps} describing the temporal
dynamics of periodically forced self-oscillators. This analogy suggests
that our system should exhibit the same richness of spatial
behaviors as the circle map does in time-evolutions.
The weak forcing case is illustrated in Fig.~\ref{fig:xn_an}. As
in the case of circle maps for very weak forcing, the front
positions return map shows a very small deviation from
linearity with the given parameter values.
This approximate linearity implies that the front train wavelength
is nearly constant and the influence of the channel walls negligible.
This influence is, however, more important on the
front heights. Minima of front height are situated at the narrowest
channel sections, in concordance with
Fig.~\ref{fig:canal_recto}, while the maxima saturate for $d$
large enough.

{}Let us now derive a semi-analytical
expression for the return maps of successive wave fronts positions
and maxima heights.
In view of the results for the weak
forcing case we assume that the front velocity in the undulated
channel at a position where the local width is $w$
adapts quasi-adiabatically to the velocity $c(w)$ (Fig.~\ref{fig:canal_recto})
corresponding to a uniform channel of the same width. Thus, the velocity of
the $n$-th front is
\begin{equation}
\dot x_n(t) = c(w(x_n))
\label{mapvel}
\end{equation}
\noindent
In our
channel $w(x)=w_0-w_1\cos(kx)$ with $w_0=s+d$ and
$w_1=d$. In order to proceed analytically
an approximation for $c(w)$
should be introduced. For $d$ small the
width variation is also small and
$c(w)$ can be replaced by a linear fit $a+bw$ of an apropriate range
of data in Fig.~\ref{fig:canal_recto}. Hence, $c(w(x_n))\approx
c_0-c_1\cos(kx_n)$ where $c_0=a+bw_0$ and $c_1=b w_1$.
Eq.~(\ref{mapvel}) can now be integrated during one period $T$
of the front generator, to obtain:
\begin{equation}
\frac{2}{k \sqrt{c_0^2-c_1^2}}
\left\lbrack\arctan f(x_n)
\right \rbrack _{x_n(0)}^{x_{n+1}(0)}=T
\label{resolcintegral}
\end{equation}
with
\begin{equation}
z_n=f(x_n)=\sqrt{\frac{c_0+c_1}{c_0-c_1}}\tan\left(\frac{kx_n}{2}\right)
\label{f}
\end{equation}
Here we have used the observed time periodicity of the wave train to
write $x_n(T)=x_{n+1}(0)$, a crucial step to convert the
time-differential equation (\ref{mapvel}) into a map for space
positions. Defining $\varphi=\arctan z$ and $\bigtriangleup
=0.5\,kT\sqrt{c_0^2-c_1^2}$ we have
$\bigtriangleup=\varphi_{n+1}-\varphi_n$, and the return map for the
variable $z$ is
\begin{equation}
z_{n+1}=g(z_n)=\frac{z_n+\tan \bigtriangleup}{1-\tan
\bigtriangleup \cdot z_n}\ .
\label{ache}
\end{equation}
In terms of the front position $x$ we finally have:
\begin{equation}
x_{n+1}=f^{-1}\left(g(f(x_n))\right)
\label{mapx}
\end{equation}

For the maximum height of the wave fronts, the same
adiabaticity assumption leads to
$a_n=h\left(w_0-w_1\cos(k x_n)\right)$, with $h(w)$ being the maximum
height of the fronts in a straight channel of width $w$.
We can go one step further towards qualitatively describe
the observed positional mismatch between the minimal-height fronts and
the narrowest channel sections by considering a short
adaptation time $\tau_a$ of the front characteristics to the local width:
\begin{equation}
\dot a_n(t) = \frac{1}{\tau_a}\left (h(w(x_n))-a_n \right)
\label{ecualt}
\end{equation}
As above, by linearly fitting the data from
Fig.~\ref{fig:canal_recto} in the range $(s,s+2d)$
we have $h(w)\approx a'+b'w$ for small $d$. Then,
$h(w(x)) \approx h_0-h_1\cos kx$, with
$h_0=a'+b'w_0$, and $h_1=b'w_1$. Integrating Eq.~(\ref{ecualt})
for small $d$ and $kc_0\tau_a
\ll 1$ so that we can set $x_n(t)=x_n(0)+c_0 t+
{\cal O} (d)$, we get a relationship linking the
wavefront heights and positions,
\begin{equation}
a_{n+1}=h_0-\frac{h_1}{\sqrt{1+\tau_a^2
k^2v_0^2}}\sin\left(k(x_n+v_0T)+\theta\right)\ .
\label{mapalt}
\end{equation}
\noindent
Here $\theta=\arccos(1/\sqrt{1+(\tau _a v_0k)^2})$ describes the
displacement of the minimal heights from the narrowest sections.
\begin{center}
\begin{figure}
\epsfig{file=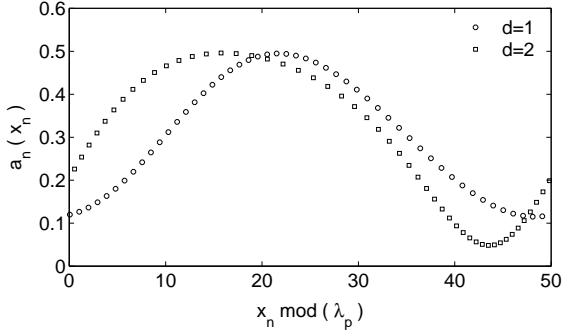, width=7.5cm }
\vspace{0.1in}
\caption[]{Change in the shape of $a_n(x_n)$ from
Eq.~(\ref{mapalt}) by increasing $d$ beyond the weak forcing
limit. The qualitative trend agrees with
Fig.~\ref{fig:TresxSete}.}
\label{fig:num_teor}
\end{figure}
\end{center}
Since the derivation of Eqs.~(\ref{mapx}) and (\ref{mapalt})
is formally valid only in the weak forcing limit we first
contrast the theory against the numerical data in
Fig.~\ref{fig:xn_an} for $d=0.5$, to confirm the good
agreement\cite{fitting}. More detailed numerical explorations
reasure us that both adiabaticity and small $d$ approximations
are justified and that the small
deviations in Fig.~\ref{fig:xn_an} are only due to the linear
approximation. Moreover, a systematic $d$-expansion in
Eq.~(\ref{mapx}) would lead precisely to a circle map supporting
the observation that this model is relevant to the
description of our boundary-induced patterns
in a given limit. Finally, while the agreement between
the theory and the numerics is bound to worsen as forcing increases, 
the theory still describes well
the pattern features in the strong forcing regime. For instance,
Fig.~\ref{fig:num_teor} shows how the maxima and minima of
$a_n(x_n)$ shift as $d$ is increased, in good qualitative agreement
with the corresponding case of Fig.~\ref{fig:TresxSete}.

In summary, we have shown that boundary conditions in domains with
the form of undulated channels may induce nontrivial longitudinal spatial
configurations of propagating
trains of excitation fronts
generated by a local time-periodic stimulation in simple 
excitable media. In particular, stroboscopically frozen
quasi-periodic arrays of fronts were found. These structures
were described in terms of spatial return maps that
are very similar to the circle maps
whose iteration describe the temporal dynamics of forced oscillators.
This similarity allows one to speculate about the existence of
even more complex configurations representing the spatial
realizations of the chaotic regimes of these maps.
The phenomenon reported here should be experimentally observable in the
photo-sensitive Belusov-Zhabotinsky reaction with proper lighting
conditions at the boundaries. Work along this line is currently
in progress.

Finantial support from DGES projects PB94-1167, PB97-0540 and 
PB97-0141-C01-01 is greatly acknowledged.

\end{multicols}
\end{document}